\documentclass[prx, reprint, amsmath, amssymb, superscriptaddress]{revtex4-2} 
\usepackage{graphicx}
\usepackage{multirow, dcolumn}
\usepackage{bm, physics}
\usepackage{url}
\usepackage{enumitem} \setlist{nosep}
\usepackage{xcolor}
\usepackage{tabularx}
\usepackage{booktabs}
\usepackage{hyperref}
\hypersetup{
    breaklinks,
    colorlinks,
    citecolor=blue,
    urlcolor=blue,
    linkcolor=purple
}

\begin{document}

\title{General framework for E(3)-equivariant neural network representation of density functional theory Hamiltonian}

\newcommand{\thuphy}{State Key Laboratory of Low Dimensional Quantum Physics and Department of Physics, Tsinghua University, Beijing, 100084, China}
\newcommand{\pkuphy}{School of Physics, Peking University, Beijing 100871, China}
\newcommand{\thuias}{Institute for Advanced Study, Tsinghua University, Beijing 100084, China}
\newcommand{\tencent}{Tencent Quantum Laboratory, Tencent, Shenzhen, Guangdong 518057, China}
\newcommand{\fscqi}{Frontier Science Center for Quantum Information, Beijing, China}
\newcommand{\baqis}{Beijing Academy of Quantum Information Sciences, Beijing 100193, China}
\newcommand{\riken}{RIKEN Center for Emergent Matter Science (CEMS), Wako, Saitama 351-0198, Japan}

\affiliation{\thuphy}
\affiliation{\pkuphy}
\affiliation{\tencent}
\affiliation{\fscqi}
\affiliation{\thuias}
\affiliation{\baqis}
\affiliation{\riken}

\author{Xiaoxun Gong}
\affiliation{\thuphy}
\affiliation{\pkuphy}

\author{He Li}
\affiliation{\thuphy}
\affiliation{\thuias}

\author{Nianlong Zou}
\affiliation{\thuphy}

\author{Runzhang Xu}
\affiliation{\thuphy}

\author{Wenhui Duan}
\email{duanw@tsinghua.edu.cn}
\affiliation{\thuphy}
\affiliation{\tencent}
\affiliation{\fscqi}
\affiliation{\thuias}
\affiliation{\baqis}

\author{Yong Xu}
\email{yongxu@mail.tsinghua.edu.cn}
\affiliation{\thuphy}
\affiliation{\tencent}
\affiliation{\fscqi}
\affiliation{\riken}

\begin{abstract}
Combination of deep learning and \textit{ab initio} calculation has shown great promise in revolutionizing future scientific research, but how to design neural network models incorporating \textit{a priori} knowledge and symmetry requirements is a key challenging subject. Here we propose an E(3)-equivariant deep-learning framework to represent density functional theory (DFT) Hamiltonian as a function of material structure, which can naturally preserve the Euclidean symmetry even in the presence of spin-orbit coupling. Our DeepH-E3 method enables very efficient electronic-structure calculation at \textit{ab initio} accuracy by learning from DFT data of small-sized structures, making routine study of large-scale supercells ($> 10^4$ atoms) feasible. Remarkably, the method can reach sub-meV prediction accuracy at high training efficiency, showing state-of-the-art performance in our experiments. The work is not only of general significance to deep-learning method development, but also creates new opportunities for materials research, such as building Moir\'e-twisted material database.
\end{abstract}

\maketitle

\section{INTRODUCTION}

\begin{figure*}[t]
    \centering
    \includegraphics[width=0.8\linewidth]{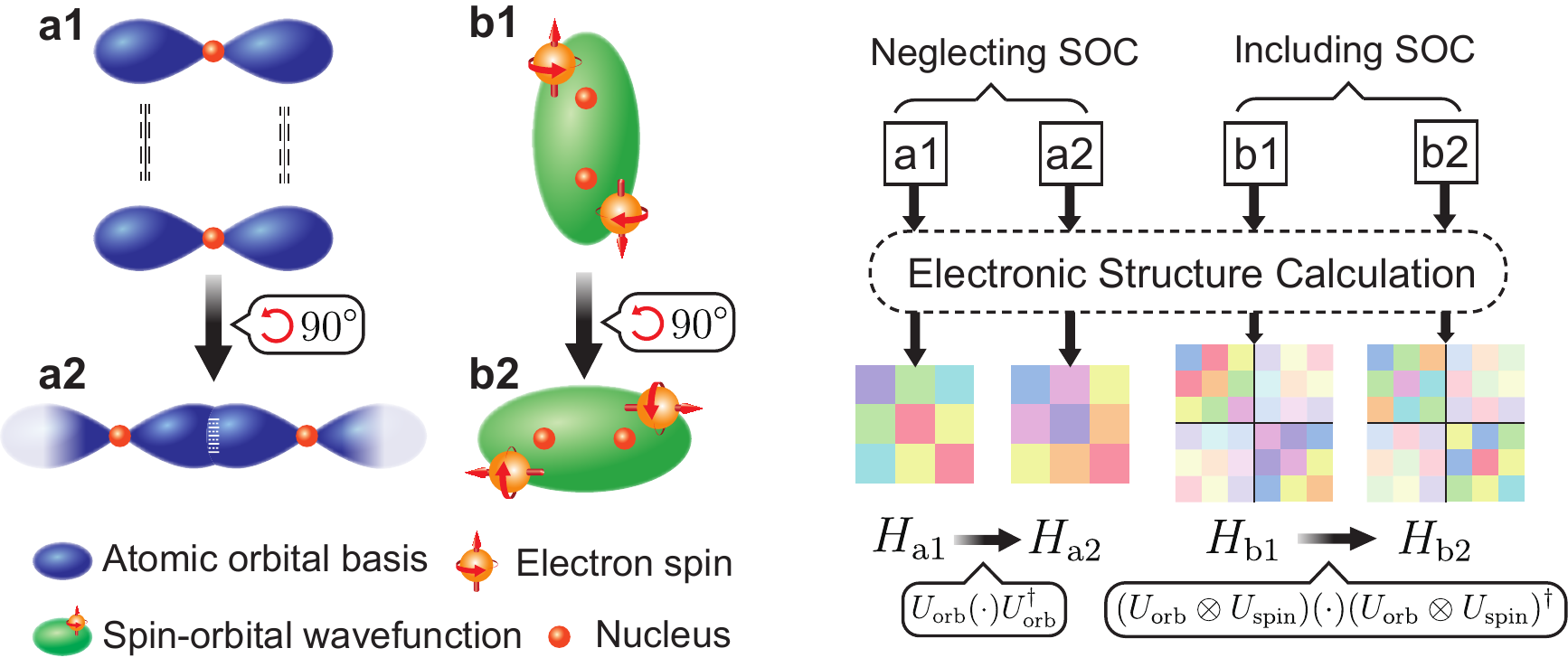}
    \caption{Equivarience in electronic structure calculations. Schematic wavefunctions and Hamiltonian matrices are shown for the systems neglecting or including spin-orbit coupling (SOC). Structures a1 and a2 are related to each other by a 90$^\circ$ rotation, whose hopping parameters (i.e., Hamiltonian matrix elements) between $p_x$ orbitals are related by a unitary transformation. This equivariant property of Hamiltonian must be preserved in all electronic structure calculations. When the SOC are taken into account, the spin and orbital degrees of freedom are coupled and must transform together under global rotations, as shown for structures b1 and b2. } 
    \label{fig:1-equivariance.pdf}
\end{figure*}

It has been well recognized that deep learning methods could offer a potential solution to the accuracy-efficiency dilemma of \textit{ab initio} material calculations. Deep-learning potential~\cite{behler2007, zhang2018} and a series of other neural network models~\cite{ani2017, schnet2018, spookynet2021, gemnet2021, nequip2022} are capable of predicting the total energies and atomic forces of given material structures, enabling molecular dynamics simulation at large length and time scales. The paradigm has been used for deep-learning research of various kinds of physical and chemical properties~\cite{duvenaud2015, gilmer2017, chandrasekaran2019, schnorb2019, dimenet2020, cormorant2019, painn2021, unite2021, jorgensen2018, cgcnn2018, physnet2019,  su2022}. Remarkably, a deep neural network representation of density functional theory (DFT) Hamiltonian (named DeepH) was developed by employing the locality of electronic matter, localized basis, and local coordinate transformation~\cite{deeph2022}. By the DeepH approach the computationally demanding self-consistent field iterations could be bypassed and all the electron-related physical quantities in the single-particle picture can in principle be derived very efficiently. This opens opportunities for the electronic-structure calculation of large-scale material systems. 

Introducing physical insights and \textit{a priori} knowledge into neural networks is of crucial importance to the deep-learning approaches. Specifically, the deep-learning potential takes advantage of the invariance of the total energy under  rotation, translation and spatial inversion as well as permutation of atoms. For DeepH, the property that the Hamiltonian matrix changes covariantly (i.e. equivariantly) under rotation or gauge transformations should be preserved by the neural network model for efficient learning and accurate prediction (Fig.~\ref{fig:1-equivariance.pdf}). A strategy is developed to apply local coordinate transformation which changes the rotation covariant problem into an invariant one and thus the transformed Hamiltonian matrices can be learned flexibly via rotation-invariant neural networks~\cite{deeph2022}. Nevertheless, the large amount of local coordinate information seriously increases the computational load, and the model performance depends critically on a proper selection of local coordinates, which relies on human intuition and is not easy to optimize. Alternatively, one may get rid of the local coordinate transformation by applying the equivariant neural network (ENN)~\cite{cohen2016, tensorfield2018, se3cnn2018, cgnets2018}. The key innovation of ENN is that all the internal features transform under the same symmery group with the input, thus the symmetry requirements are explicitly treated and exactly satisfied, as shown by a series of neural network models for various material properties~\cite{cormorant2019, painn2021, unite2021, gemnet2021, nequip2022}, including PhiSNet~\cite{phisnet2021} for predicting the Hamiltonian of molecules with fixed system size. However, the key capability of DeepH that learns from DFT results on small-sized material systems and predicts the electronic structures of much larger ones has not been demonstrated by ENN models. More critically, the existing ENN models have neglected the equivariance in the spin degrees of freedom, although the electronic spin and spin-orbit coupling (SOC) play a key role in modern condensed matter physics and materials science. With SOC, one should take care of the spin-orbital Hamiltonian, whose spin and orbital degrees of freedom are coupled and transform together under change of coordinate system or basis set, as illustrated in Fig.~\ref{fig:1-equivariance.pdf}. This would raise critical difficulties in designing ENN models due to a fundamental change of symmetry group. In this context, the incorporation of ENN models into DeepH is essential but remains elusive.

In this work, we propose DeepH-E3, a universal E(3)-equivariant deep-learning framework to represent the spin-orbital DFT Hamiltonian $\hat H_\text{DFT}$ as a function of atomic structure $\{\mathcal R\}$ by neural networks, which enables very efficient electronic structure calculations of large-scale materials at \textit{ab initio} accuracy. A general theoretical basis is developed to explicitly incorporate covariance transformation requirements of $\{\mathcal R\}\mapsto\hat H_\text{DFT}$ into neural network models that can properly take the electronic spin and SOC into account, and a code implementation of DeepH-E3 based on message passing neural network is also presented. Since the principle of covariance is automatically satisfied, efficient learning and accurate prediction become feasible via the DeepH-E3 method. Our systematic experiments demonstrate state-of-the-art performance of DeepH-E3, which shows sub-meV accuracy in predicting DFT Hamiltonian. The method works well for various kinds of material systems, such as magic-angle twisted bilayer graphene or twisted van der Waals materials in general, and the computational costs are reduced by several orders of magnitude compared to direct DFT calculations. 
Benefiting from the high efficiency and accuracy as well as the good transferability, there could be promising applications of DeepH-E3 in electronic structure calculations. Also we expect that the proposed neural-network framework can be generally applied to develop deep-learning \textit{ab initio} methods and that the interdisciplinary developments would eventually revolutionize future materials research.  

\section{Realization of equivariance}

\begin{figure*}[t]
    \centering
    \includegraphics[width=0.8\linewidth]{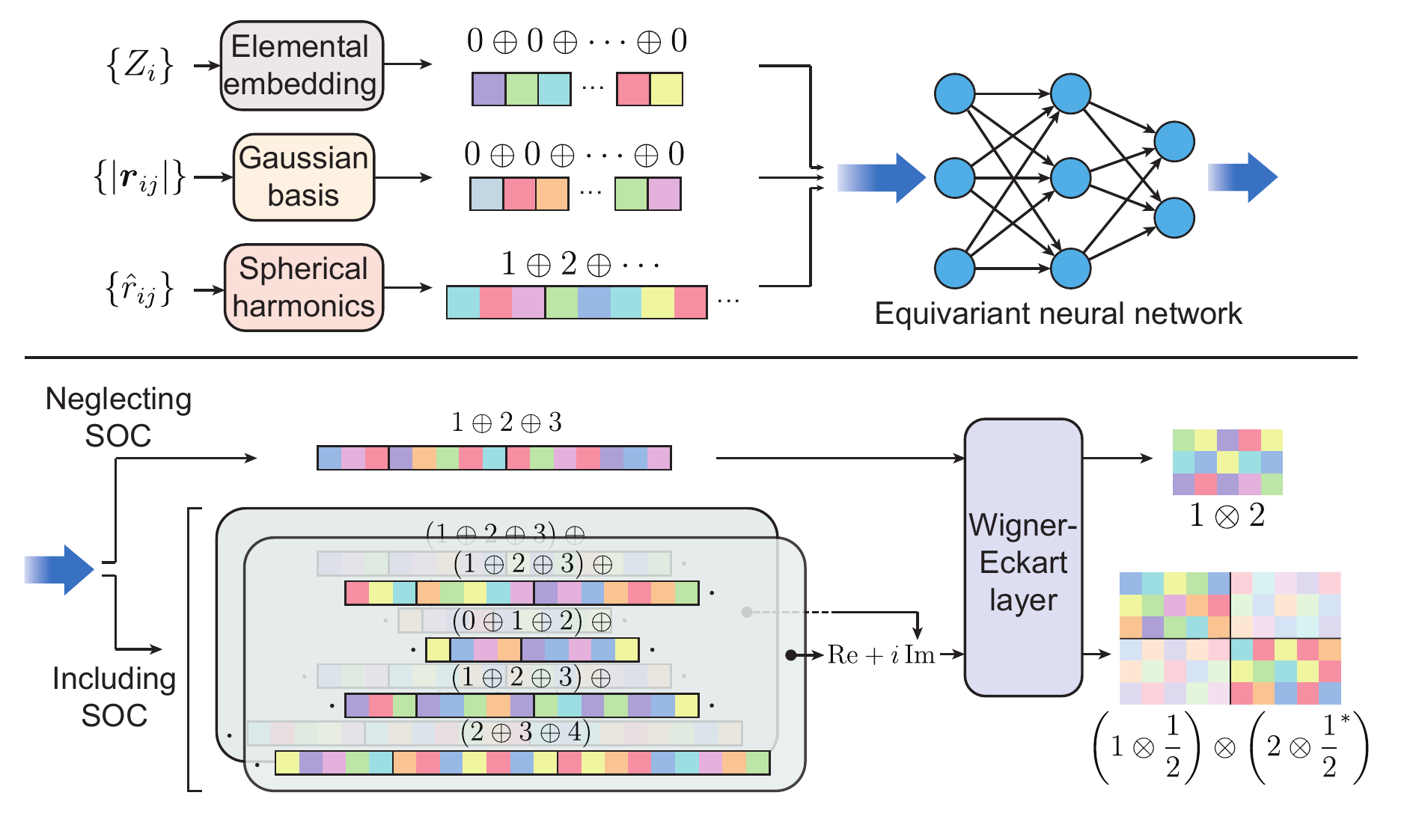}
    \caption{Method of constructing an equivariant mapping $\{\mathcal{R}\} \mapsto \hat{H}_{\mathrm{DFT}}$. Take the Hamiltonian matrix between $l=1$ and $l=2$ orbitals for example. The atomic numbers $Z_i$ and interatomic distances $|\boldsymbol r_{ij}|$ are used to construct the $l=0$ vectors, and the unit vectors of relative positions $\hat r_{ij}$ are used to construct vectors of $l=1,2,\dots$. These vectors are passed to the equivariant neural network. If neglecting spin-orbit coupling (SOC), the output vectors of the neural network are converted to the Hamiltonian using the rule $1\oplus 2\oplus 3=1\otimes 2$ via the Wigner-Eckart layer. If including SOC, the output consists of two sets of real vectors which are combined to form complex-valued vectors. These vectors are converted to the spin-orbital DFT Hamiltonian according to a different rule $(1 \oplus 2 \oplus 3) \oplus(0 \oplus 1 \oplus 2) \oplus(1 \oplus 2 \oplus 3) \oplus(2 \oplus 3 \oplus 4) = \left(1 \otimes \frac{1}{2}\right) \otimes\left(2 \otimes \frac{1}{2}^*\right)$.}
    \label{fig:2-wigner-eckart.pdf}
\end{figure*}

It has long been established as one of the fundamental principles of physics that all physical quantities must transform equivariantly between reference frames. Formally, a mapping $f: X\to Y$ is equivariant for vector spaces $X$ and $Y$ with respect to group $G$ if $D_Y(g)\circ f = f\circ D_X(g),\forall g\in G$, where $D_X,D_Y$ are representations of group $G$ over vector spaces $X,Y$, respectively. The problem considered in this work is the equivariance of a mapping from the material structure $\{\mathcal R\}$ including atom types and positions to the DFT Hamiltonian $\hat H_\text{DFT}$ with respect to the E(3) group. The E(3) group is the Euclidean group in three-dimensional (3D) space which contains translations, rotations and inversion. Translation symmetry is manifest since we only work on the relative positions between atoms, not their absolute positions. Rotations of coordinates introduce non-trivial transformations which should be carefully investigated. Suppose the same point in space is specified in two coordinate systems by $\boldsymbol r$ and $\boldsymbol r'$. If the coordinate systems are related to each other by a rotation, the transformation rule between the coordinates of the point is $\boldsymbol r'=\mathbf{R} \boldsymbol{r}$, where $\mathbf R$ is a $3\times 3$ orthogonal matrix.

In order to take advantage of the nearsightedness of electronic matter~\cite{Kohn2005}, the Hamiltonian operator is expressed in the picture of localized atomic orbital basis. The basis is separated into radial and angular parts, having the form $ \phi_{i \alpha}(\boldsymbol{r})=R_{i p l}(r) Y_{l m}(\hat{r}) $. Here $i$ is the site index, $\alpha\equiv(plm)$, where $p$ is the multiplicity index, $Y_{lm}$ is the spherical harmonics having angular momentum quantum number $l$ and magnetic quantum number $m$, $r \equiv |\boldsymbol r - \boldsymbol r_i|$ and $\hat{r} \equiv\left(\boldsymbol{r}-\boldsymbol{r}_{i}\right) /\left|\boldsymbol{r}-\boldsymbol{r}_{i}\right|$. The transformation rule for the Hamiltonian matrix between the two coordinate systems described above is
\begin{multline}
    \left(H_{i p_{1}, j p_{2}}^{\prime}\right)_{m_{1} m_{2}}^{l_{1} l_{2}}=\sum_{m_{1}^{\prime}=-l_{1}}^{l_{1}} \sum_{m_{2}^{\prime}=-l_{2}}^{l_{2}} \\
    D_{m_{1} m_{1}^{\prime}}^{l_{1}}(\mathbf{R}) D_{m_{2} m_{2}^{\prime}}^{l_{2}}(\mathbf{R})^{*}\left(H_{i p_{1}, j p_{2}}\right)_{m_{1}^{\prime} m_{2}^{\prime}}^{l_{1} l_{2}} ,
    \label{eq:transform_H}
\end{multline}
where $D^l_{mm'}(\mathbf{R})$ is the Wigner D-matrix. The equivariance of the mapping $ \{\mathcal{R}\} \mapsto \hat{H}_{\mathrm{DFT}} $ requires that, if the change of coordinates causes the positions of the atoms to transform, the corresponding Hamiltonian matrix must transform covariantly according to Eq.~\eqref{eq:transform_H}. If we further consider the spin degrees of freedom, the transformation rule for the Hamiltonian becomes
\begin{multline}
    \left(H_{i p_{1}, j p_{2}}^{\prime}\right)_{m_{1} \sigma_{1} m_{2} \sigma_{2}}^{l_{1}\frac{1}{2} l_{2} \frac{1}{2}}= \\ \sum_{m_{1}^{\prime}=-l_{1}}^{l_{1}} \sum_{m_{2}^{\prime}=-l_{2}}^{l_{2}} \sum_{\sigma_{1}^{\prime}=\uparrow, \downarrow} \sum_{\sigma_{2}^{\prime}=\uparrow, \downarrow}
    D_{m_{1} m_{1}^{\prime}}^{l_{1}}(\mathbf{R}) D_{\sigma_{1} \sigma_{1}^{\prime}}^{\frac{1}{2}}(\mathbf{R}) \\
    D_{m_{2} m_{2}^{\prime}}^{l_{2}}(\mathbf{R})^{*} D_{\sigma_{2} \sigma_{2}^{\prime}}^{\frac{1}{2}}(\mathbf{R})^{*}\left(H_{i p_{1}, j p_{2}}\right)_{m_{1}^{\prime} \sigma_{1}^{\prime} m_{2}^{\prime} \sigma_{2}^{\prime}}^{l_{1}\frac{1}{2} l_{2} \frac{1}{2}}, \label{eq:transform-H-spin}
\end{multline}
where $\sigma_1,\sigma_2$ are the spin indices (spin up or down). 

ENN is applied to construct the mapping $ \{\mathcal{R}\} \mapsto \hat{H}_{\mathrm{DFT}} $ in order to preserve equivariance. The input, output and internal features of ENNs all belong to a special set of vectors which have the form $\boldsymbol x_l=(x_{l,l},\dots, x_{l,-l})$ and transform according to the following rule:
\begin{equation}
    x_{l m}^{\prime}=\sum_{m^{\prime}=-l}^{l} D_{m m^{\prime}}^{l}(\mathbf{R}) x_{l m}.  \label{eq:transform-vector}
\end{equation}
This vector is said to carry the irreducible representation of the SO(3) group of dimension $2l+1$. If the input vectors are transformed according to Eq.~\eqref{eq:transform-vector}, then all the internal features and the output vectors of the ENN will also be transformed accordingly. Under this constraint, the ENN incorporates learnable parameters in order to model equivariant relationships between inputs and outputs.

The method of constructing the equivariant mapping $ \{\mathcal{R}\} \mapsto \hat{H}_{\mathrm{DFT}} $ is illustrated in Fig.~\ref{fig:2-wigner-eckart.pdf}. The atomic numbers $Z_i$ and interatomic distances $|\boldsymbol r_{ij}|$ are used to construct the $l=0$ input vectors (scalars). Spherical harmonics acting on the unit vectors of relative positions $\hat r_{ij}$ constitute input vectors of $l=1,2,\dots$. The output vectors of the ENN are passed through the Wigner-Eckart layer before representing the final Hamiltonian. This layer exploits the essential concept of the Wigner-Eckart theorem:
\begin{equation}
    l_1 \otimes l_2 = |l_1-l_2|\oplus \cdots \oplus (l_1+l_2). \label{eq:equivalence-representation}
\end{equation}
``$\oplus$" and ``$\otimes$'' signs stand for direct sum and tensor product of representations, respectively. ``=" denotes equivalence of representations, i.e., they differ from each other by a change of basis. The coefficients in the change of basis is exactly the celebrated Clebsch-Gordan coefficients. The representation $l_1\otimes l_2$ is carried by the tensor $\boldsymbol x_{l_1 l_2}$, which transforms according to the rule 
\begin{multline}
    x_{l_{1} l_{2} m_{1} m_{2}}^{\prime}=\\
    \sum_{m_{1}^{\prime}=-l_{1}}^{l_{1}} \sum_{m_{2}^{\prime}=-l_{2}}^{l_{2}}
    D_{m_{1} m_{1}^{\prime}}^{l_{1}}(\mathbf R) D_{m_{2} m_{2}^{\prime}}^{l_{2}}(\mathbf R) x_{l_{1} l_{2} m_{1}^{\prime} m_{2}^{\prime}}. \label{eq:transform-coupled}
\end{multline}
Notice that Eq.~\eqref{eq:transform-coupled} has the same form as Eq.~\eqref{eq:transform_H}, so the tensor $\boldsymbol x_{l_1 l_2}$ can exactly represent the output Hamiltonian satisfying the equivariant requirements.

The construction of the spin-orbital DFT Hamiltonian is a far more complicated issue. Electron spin has angular momentum $l=1/2$, so it seems that tedious coding and debugging are unavoidable because we have to introduce complex-valued half-integer representations into the neural network which typically only supports real-valued integer representations for the time being. Furthermore, a $2\pi$ rotation brings a vector in 3D space to itself, but introduces a factor -1 to the spin-1/2 vector. This means that any mapping from 3D input vectors to $l=1/2$ output vectors will be discontinued and cannot be modeled by neural networks, which poses a serious threat to our approach since we only have 3D vectors as input to the neural network (Fig.~\ref{fig:2-wigner-eckart.pdf}). 

Fortunately, we observe that $l=1/2$ appearing in the DFT Hamiltonian does not necessarily mean that half-integer representations must be inserted everywhere into the neural network. In fact, they can be restricted to the final output layer, as soon as we employ the transformation rule:
\begin{equation}
    \left(l_{1} \otimes \frac{1}{2}\right) \otimes\left(l_{2} \otimes \frac{1}{2}\right)=\left(l_{1} \otimes l_{2}\right) \otimes(0 \oplus 1).
    \label{eq:eliminate-spin}
\end{equation}
There is no half-integer representation on the right hand side, thus it can be further decomposed into integer representations by repeatedly applying Eq.~\eqref{eq:equivalence-representation}.

Another problem is associated with the introduction of complex numbers. Generally, the spin-orbital Hamiltonian matrix elements have complex values, and the ENN cannot simply predict its real and imaginary parts separately because this will violate equivariance. Ordinary neural networks of complex numbers are mostly still under their experimental and developmental stage, so the use of complex-valued ENN is practically difficult, if not impossible. Nevertheless, we have discovered a way to sidestep this problem. Under the bases which are eigenvectors of the time-reversal operator, the D-matrices of integer $l$ will become purely real. Consequently, for a vector with integer $l$ under that basis, its complex and real part will never mingle with each other when the vector is multiplied by a real transformation matrix. Then one complex vector can be technically treated as two real vectors while preserving equivariance. Note that this is not true for half-integer representations, for that we must add up the real and imaginary parts before the integer representations are converted to half-integer representations in the Wigner-Eckart layer (Fig.~\ref{fig:2-wigner-eckart.pdf}).

Yet another subtle issue arises in Eq.~\eqref{eq:transform-H-spin}. It is not exactly the same as Eq.~\eqref{eq:transform-coupled} in that two of the D-matrices in the former equation are taken complex conjugate, but those in the latter are not. In fact, instead of constructing a vector with representation $\left(l_{1} \otimes \frac{1}{2}\right) \otimes\left(l_{2} \otimes \frac{1}{2}\right)$, we must construct $\left(l_{1} \otimes \frac{1}{2}\right) \otimes\left(l_{2}^* \otimes \frac{1}{2}^*\right)$ to represent the spin-orbital Hamiltonian described in Eq.~\eqref{eq:transform-H-spin}. Here $l^*$ denotes the representation whose representation matrix is replaced by its complex conjugate. This is not a problem for integer $l$, but is critical for $l=1/2$. If not treated properly, the overall equivariance will be violated. In order to solve this problem, we first notice that the representation $l^*$ is still a representation of the SU(2) group with dimension $2l+1$. In fact, it is guaranteed to be equivalent with the representation $l$ without complex conjugate. In other words, there must exist a unitary matrix $\mathbf P^l$ for each integer or half-integer $l$ satisfying 
\begin{equation}
    \mathbf{D}^{l}(g)^{*}=\mathbf{P}^l \mathbf{D}^{l}(g) (\mathbf{P}^l)^{\dagger},\forall g\in\text{SU(2)} .\label{eq:transform-cc}
\end{equation}
Indeed, this is guaranteed by the fact that the quantum rotation operator $\hat U(g)$ commutes with the time-reversal operator $\mathcal T$: $\langle l m|\hat U(g)| l m'\rangle=\langle l m|\mathcal{T}^{\dagger} \hat U(g) \mathcal{T}| l m'\rangle
    =(-1)^{m-m'}\langle l,-m|\hat U(g)| l,-m'\rangle^{*}$. The matrix $\mathbf P$ in Eq.~\eqref{eq:transform-cc} is thus given by
\begin{equation}
     P^{l}_{mm'}=(-1)^{l-m}\delta_{m,-m'}.\label{eq:transform-cc-p}
\end{equation}
Therefore, we only need to apply a change of basis to convert a vector carrying representation $l$ to a vector carrying $l^*$. Notice that this property holds even for material systems without time reversal symmetry.

The workflow of constructing the DFT Hamiltonian is summarized and illustrated in Fig.~\ref{fig:2-wigner-eckart.pdf}. In order to construct a Hamiltonian with SOC, the output vectors from the ENN are first separated into two real components, then combined together into complex vectors and passed to the Wigner-Eckart layer. The Wigner-Eckart layer uses the rules in Eq.~\eqref{eq:equivalence-representation} and Eq.~\eqref{eq:eliminate-spin} to convert these vectors to tensors of the form in Eq.~\eqref{eq:transform-coupled}, except that the tensors here have rank 4 instead of 2. After that, the last spin index is converted to the complex conjugate counterpart by the change-of-basis using Eq.~\eqref{eq:transform-cc-p} for $l=1/2$. The output tensors follow the same transformation rule under coordinate rotation as the DFT Hamiltonian in Eq.~\eqref{eq:transform-H-spin}, and thus could be used to represent the DFT Hamiltonian matrix.

Finally, we discuss two remaining issues. To include parity, we will consider $\text{E}(3)=\text{SE}(3)\otimes \{E,I\}$, where $E$ is the identity and $I$ is the spatial inversion. Under a coordinate transform, the vector is multiplied by -1 if it has odd parity and the coordinate transform involves spatial inversion. The parity of the Hamiltonian is determined by $(-1)^{l_1+l_2}$. In addition, there is a possible ambiguity in Eq.~\eqref{eq:transform-H-spin}, since the mapping from a classical rotation $\mathbf R$ to a quantum rotation $\mathbf D^{\frac{1}{2}}$ is not single-valued. However, the possible factor $-1$ will always be canceled between the two D-matrices in that equation, which eliminates the potential problem.

\begin{figure*}[t]
    \centering
    \includegraphics[width=0.93\linewidth]{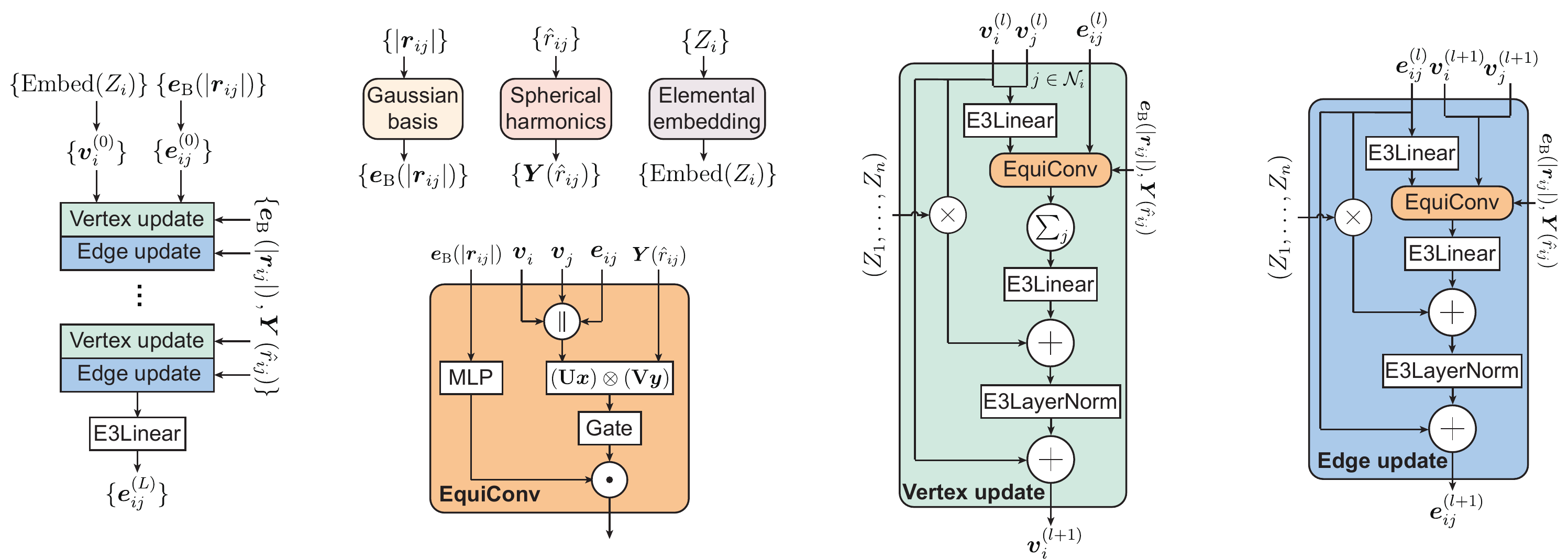}
    \caption{Neural network architecture of DeepH-E3. Elemental embeddings and Gaussian expansions serve as initial vertex and edge features, respectively. The vertex and edge features are updated $L$ times by update blocks, which encode the interatomic distances and directional information through equivariant convolutions. The ``$\cdot$'' sign stands for channel-wise multiplication and $||$ for vector concatenation. $\mathcal N_i$ is the neighborhood of vertex $i$. The final edge vectors $\{\boldsymbol e_{ij}^{(L)}\}$ is passed into the Wigner-Eckart layer depicted in Fig.~\ref{fig:2-wigner-eckart.pdf} to represent the DFT Hamiltonian.}
    \label{fig:3-net.pdf}
\end{figure*}

\section{Neural network architecture of D\lowercase{eep}H-E3}

Here we present the neural network architecture of the DeepH-E3 method. The general structure is based on the message passing neural network~\cite{gilmer2017, mpnn2018} that has been widely used in materials research~\cite{jorgensen2018, cgcnn2018, physnet2019, painn2021, unite2021, gemnet2021, nequip2022, su2022, deeph2022}. The material structure is represented by a graph, where each atom is associated with a vertex (or node). Edges are connected between atom pairs with nonzero inter-site hopping, and self-loop edges are included to describe intra-site coupling. Every vertex $i$ is associated with a feature $\boldsymbol v_i$ and every edge $ij$ with $\boldsymbol e_{ij}$. These features are composed of several vectors defined in Eq.~\eqref{eq:transform-vector}. The initial feature $\boldsymbol v_i^{(0)}$ of vertex $i$ is the trainable embedding of the atomic number $Z_i$, and the initial $\boldsymbol e_{ij}$ is the interatomic distance $|\boldsymbol r_{ij}|$ expanded using the Gaussian basis $\boldsymbol e_\text{B}(|\boldsymbol r_{ij}|)$ as defined in Eq.~\eqref{eq:gaussian-basis}. The features of vertices and edges are iteratively updated using features of their neighborhood as incoming messages. Finally, the final edge feature $\boldsymbol  e_{ij}$ is passed through a linear layer and used to construct the Hamiltonian matrix block $H_{ij}$ between atoms $i$ and $j$, as illustrated in Fig.~\ref{fig:2-wigner-eckart.pdf}. It is worth mentioning that, under the message passing scheme, the output Hamiltonian is only influenced by the information of its neighborhood environment. The nearsightedness property~\cite{Kohn2005} ensures efficient linear-scaling calculations as well as good generalization ability~\cite{deeph2022}.

The equivariant building blocks of the neural network are implemented using the scheme provided by Tensor-Field Networks~\cite{tensorfield2018}. The feature vectors $x^{(l)}_{cm}$ processed by these nerual network blocks are implemented as dictionaries with key $l$, an integer which is the order of representation of the SO(3) group. $c$ is the ``channel index'' ranging from $1$ to $n^{(l)}$, where $n^{(l)}$ is the number of channels at order $l$, and each channel refers to a vector defined in Eq.~\eqref{eq:transform-vector}. 

The E3Linear layer defined in Eq.~\eqref{eq:E3Linear} possesses learnable weights and biases, which is similar to linear layers in conventional neural networks, but only connects vectors of the same representation to preserve equivariance. The gate layer introduces equivariant nonlinearity as proposed in Ref.~\cite{se3cnn2018}, where nonlinearly activated $l=0$ vectors (i.e. scalars) are used as scaling factors (``gates") to the norms of $l\neq0$ vectors.

We propose a normalization scheme, E3LayerNorm, that normalizes the feature vectors using mean and variance obtained from the layer statistics while preserving equivariance:
\begin{equation}
\operatorname{E3LayerNorm}(\boldsymbol v_i)^{(l)}_{cm} = g^{(l)}_c\frac{(\boldsymbol v_i)^{(l)}_{cm}-\mu^{(l)}_m}{\sigma^{(l)}+\epsilon} +b^{(l)}_c,
\end{equation} 
where $\epsilon$ is introduced to maintain numerical stability, $g^{(l)}_c,b_c^{(l)}$ are learnable affine parameters, the mean $\mu_m^{(l)}=\frac{1}{Nn^{(l)}}\sum_{i=1}^N\sum_{c=1}^{n^{(l)}}(\boldsymbol v_i)^{(l)}_{cm}$, the variance $(\sigma^{(l)})^2=\frac{1}{Nn^{(l)}}\sum_{i=1}^N\sum_{c=1}^{n^{(l)}}\sum_{m=-l}^l\left|(\boldsymbol{v}_i)^{(l)}_{cm}-\mu_m^{(l)}\right|^2$, $N$ is the total number of vertices. Here only the E3LayerNorm for vertex update blocks is described. The corresponding E3LayerNorm for edge update blocks is similar with the mean and variance obtained from edge features instead of vertex features.

The previously discussed blocks do not include coupling between different $l$'s. This problem is resolved by the tensor product layer:
\begin{equation}
    z^{(l)}_{cm} = \sum_{l_1l_2}\sum_{c_1c_2}C^{lm}_{l_1m_1;l_2m_2}
    \left(U^{(l_1)}_{cc_1}x^{(l_1)}_{c_1m_1}\right)
    \left(V^{(l_2)}_{cc_2}y^{(l_2)}_{c_2m_2}\right),
    \label{eq:tensor-product}
\end{equation}
where $C^{l_1m_1}_{l_2m_2;l_3m_3}$ are Clebsch-Gordan coefficients, $U^{l}_{cc'},V^{l}_{cc'}$ are learnable weights. This is abbreviated as $\boldsymbol z=(\mathbf{U}\boldsymbol{x})\otimes(\mathbf{V}\boldsymbol{y})$. 

The neural network structure is illustrated in Fig.~\ref{fig:3-net.pdf}. The equivariant convolution block (EquiConv) encodes the information of an edge and the vertices connected to that edge. The core component of equivariant convolution is the tensor product (Eq.~\eqref{eq:tensor-product}) of the vertex and edge features ($\boldsymbol v_i||\boldsymbol v_j||\boldsymbol e_{ij}$) and the spherical harmonics of the edge $ij$ ($\boldsymbol Y(\hat r_{ij})$). The tensor product introduces directional information of material structure into the neural network. Propagating directional information into neural networks is important as emphasized by previous works~\cite{dimenet2020, painn2021}, which is realized in an elegant way here via the tensor product. The interatomic distance information is also encoded in neural network using the Gaussian basis expansion and fed into a fully connected neural network, whose output is multiplied element-wise to the output of gate nonlinearity.

The vertex update block aggregates information from neighboring environment. Every edge connected to the vertex of interest contributes a ``message'' generated by the equivariant convolution (EquiConv) block. All the ``messages'' are summed and normalized to update the vertex feature. This is similar for the edge update layer, except that only the output of EquiConv on edge ${ij}$ is used for updating $\boldsymbol e_{ij}$. After several updates, the final edge feature vectors will serve as the neural network output and are passed into the Wigner-Eckart layer to construct the Hamiltonian matrix blocks as illustrated in Fig.~\ref{fig:2-wigner-eckart.pdf}. More details are described in the appendix.

\section{Capability of D\lowercase{eep}H-E3}

The incorporation of global Euclidean symmetry as \textit{a priori} knowledge provided to the message-passing deep-learning framework in the DeepH-E3 model has led to its outstanding performance in terms of efficiency and accuracy. A remarkable capability of DeepH-E3 is to learn from DFT data on small structures and make predictions on varying structures of different sizes without having to perform further DFT calculations. This enables highly efficient electronic structure calculations of large-scale material systems at \textit{ab initio} accuracy. After example studies on monolayer graphene and MoS$_2$ datasets, we will first demonstrate the capability of DeepH-E3 by investigating twisted bilayer graphene (TBG), especially the well-known magic-angle TBG whose DFT calculation is important but quite challenging due to its huge Moir\'e supercell. Next, we will apply DeepH-E3 to study twisted van der Waals (vdW) materials with strong SOC, including bilayers of bismuthene, Bi$_2$Se$_3$ and Bi$_2$Te$_3$ for demonstrating the effectiveness of our equivariant approach to construct the spin-orbital DFT Hamiltonian. Finally, we will use our model to illustrate the SOC-induced topological quantum phase transition in twisted bilayer Bi$_2$Te$_3$, giving an example of exploring exotic physical properties in large-scale material systems.

\begin{figure*}[t]
    \centering
    \includegraphics[width=0.9\linewidth]{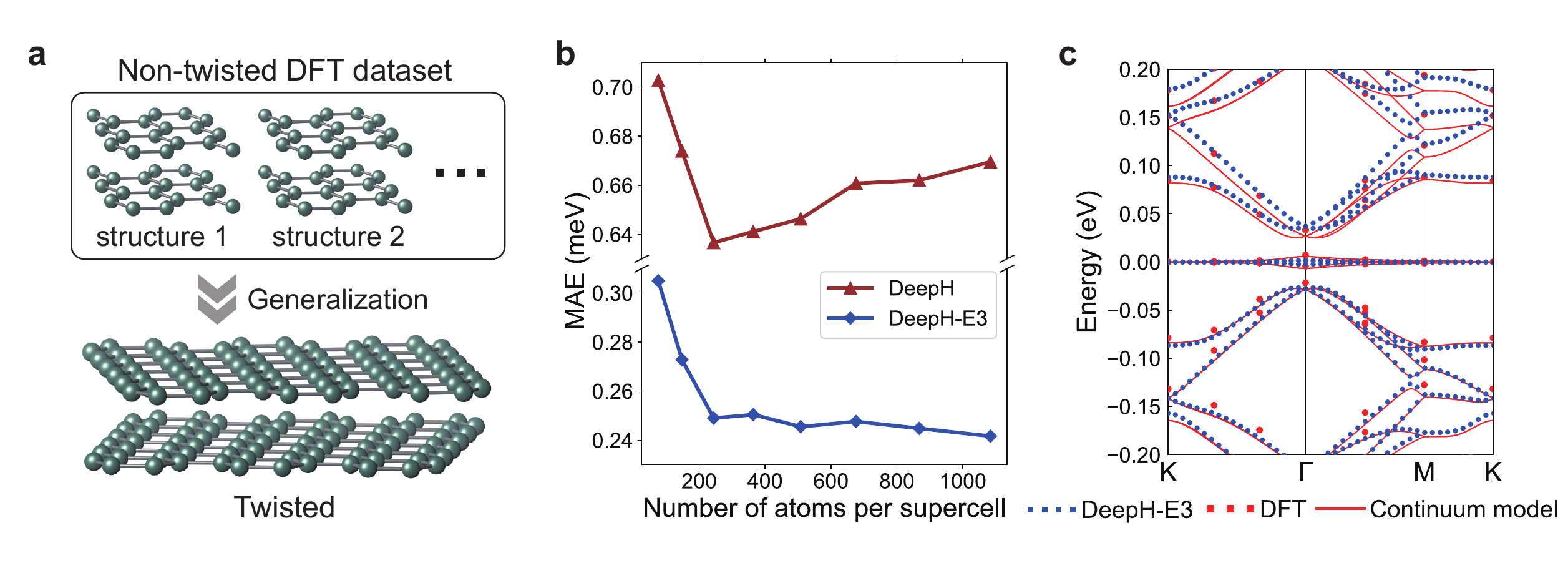}
    \caption{Application of DeepH-E3 to study twisted bilayer graphene (TBG). (\textbf{a}) Workflow of DeepH-E3. The neural network model is first trained on DFT data of small, non-twisted, randomly perturbed structures, and then generalized to study arbitrarily twisted structures without invoking DFT anymore. (\textbf{b}) Performance of DeepH-E3 vs. the original DeepH method~\cite{deeph2022} on studying TBGs of varying twist angle $\theta$. The averaged MAEs of DFT Hamiltonian are displayed for Moir\'e supercells of varied sizes. (\textbf{c}) Band structure of the magic-angle TBG ($\theta=1.08^\circ$, 11,164 atoms per supercell, structure relaxed by previous work~\cite{matbg-vasp2019}) computed by DeepH-E3, DFT and continuum model~\cite{matbg-vasp2019}. Here the DFT benchmark calculations were performed with a different code using plane-wave basis instead of atomic-like basis and different pseudopotential, which could introduce numerical differences with respect to DeepH-E3.}
    \label{fig:4.2-tbg.pdf}
\end{figure*}

Before going to large-scale materials, we first validate our method on the datasets used by Ref.~\cite{deeph2022} to benchmark DeepH-E3's performance. The datasets are comprised of DFT supercell calculation results of monolayer graphene and MoS$_2$, and different geometric configurations are sampled from \textit{ab initio} molecular dynamics. The test results are summarized in Tab.~\ref{tab:mae-graphene-mos2} and compared with those of the original DeepH method~\cite{deeph2022} which, instead of using an explicitly equivariant approach, applied the local coordinate technique in handling the covariant transformation property of the Hamiltonian. Our experiments show that the mean absolute errors (MAEs) of Hamiltonian matrix elements averaged over atom pairs are all within a fraction of a meV, which are reduced approximately by a factor of 2 or more in all prediction targets compared with DeepH. Benefiting from the high accuracy of deep-learning DFT Hamiltonian, band structures predicted by DeepH-E3 can accurately reproduce DFT results (supplementary Fig. S1).

\begin{table}[htbp]
    \centering
    \caption{MAEs of DFT Hamiltonian matrix elements averaged over atom pairs for monolayer graphene and MoS$_2$, all in units of meV. \footnote{The best result of each target is marked as \textbf{bold}. For MoS$_2$ there are four different types of atom pairs, whose MAEs are listed separately.}}
    \begin{tabular}{cccccc}
        \toprule
         & Graphene & \multicolumn{4}{c}{MoS$_2$} \\
        \cmidrule{2-6}
         & C-C & Mo-Mo & Mo-S & S-Mo & S-S \\
        \midrule
        DeepH & 2.1 & 1.3 & 1.0 & 0.8 & 0.7 \\
        DeepH-E3 & \textbf{0.40} & \textbf{0.51} & \textbf{0.46} & \textbf{0.45} & \textbf{0.37} \\
        \bottomrule
    \end{tabular}
    \label{tab:mae-graphene-mos2}
\end{table}

Our deep learning method is particularly useful for studying electronic structure of twisted vdW materials. This class of materials have attracted great interest for research and applications, since their Moir\'e super periodicity offers a new degree of freedom to tune many-body interactions and brings in emergent quantum phenomena, such as correlated states~\cite{cao2018a}, unconventional superconductivity~\cite{cao2018b} and higher-order band topology~\cite{hobt2021}. Traditionally, it is challenging to perform computationally demanding DFT calculations on large Moir\'e structures. However, this challenge could be largely overcome by DeepH-E3. One may train the neural network models by DFT data on small, non-twisted, randomly perturbed structures and predict DFT Hamiltonian of arbitrarily twisted structures bypassing DFT via deep learning, as illustrated in Fig.~\ref{fig:4.2-tbg.pdf}a. Typically, only a few hundreds of DFT training calculations are needed. This demonstrates the exceptionally high data efficiency of our method, making it even more cost-effective. 

Once the model is trained, it can be applied to study TBGs of varying twist angles. The performance is compared with that of DeepH. Test data includes DFT results for systems containing up to more than one thousand atoms per supercell. As summarized in Fig.~\ref{fig:4.2-tbg.pdf}b, DeepH-E3 significantly reduces the averaged MAEs of DFT Hamiltonian matrix elements by more than a factor of 2 as compared to DeepH, consistent with the above conclusion. Moreover, the MAEs reach ultralow values of $0.2$--$0.3$ meV and gradually decrease with increasing Moir\'e supercell size (or decreasing twist angle). This demonstrates good generalizability of DeepH-E3. The method is thus expected suitable for studying TBGs with small twist angles that are of current interest~\cite{matbg-vasp2019}.

We take the magic-angle TBG with $\theta=1.08^\circ$ and 11,164 atoms per supercell as a special example. The discoveries of novel physics relevant to flat bands in this system have triggered enormous interest in investigating twisted vdW materials. Due to the large supercell, DFT study of magic-angle TBG is a formidable task, but DeepH-E3 can routinely study such kind of material systems in a very accurate and efficient way. As shown in Fig.~\ref{fig:4.2-tbg.pdf}c, the electronic bands of magic-angle TBG with relaxed structure computed by DeepH-E3 agrees well with the published results obtained by DFT and low-energy effective continuum model~\cite{matbg-vasp2019}. The flat-bands near the Fermi level are well reproduced. Some minor discrepancies appear away from the Fermi level, which could be partially explained by the methodological difference: the benchmark work uses the plane-wave basis, whereas our work employs the atomic-like basis, and the pseudopotential used is also different.

Most remarkably, DeepH-E3 has the capability to reduce the computational cost to study these large material systems by several orders of magnitude. The DFT calculation (including structural relaxation) on magic angle TBG performed by Ref.~\cite{matbg-vasp2019} took around one month on about five thousand CPU cores. In contrast, the major computational cost of DeepH-E3 comes from the neural-network training which takes tens of GPU hours but is only required to be done once. After that, DFT Hamiltonian matrices can be constructed very efficiently via neural-network inference. The process time is on the order of minutes by one GPU for magic-angle TBG, which grows linearly with Moir\'e supercell size. Generalized eigenvalue problems are solved for 60 bands near the Fermi level to obtain the band dispersion, which only requires about 8 minutes per $\boldsymbol k$-point for magic-angle TBG using 64 CPU cores. The unprecedentedly low computational cost and high accuracy of DeepH-E3 demonstrate its potential power in resolving the accuracy-efficiency dilemma of \textit{ab initio} calculation methods, and it would be highly favorable to future scientific research.

\begin{figure*}[t]
    \centering
    \includegraphics[width=0.9\linewidth]{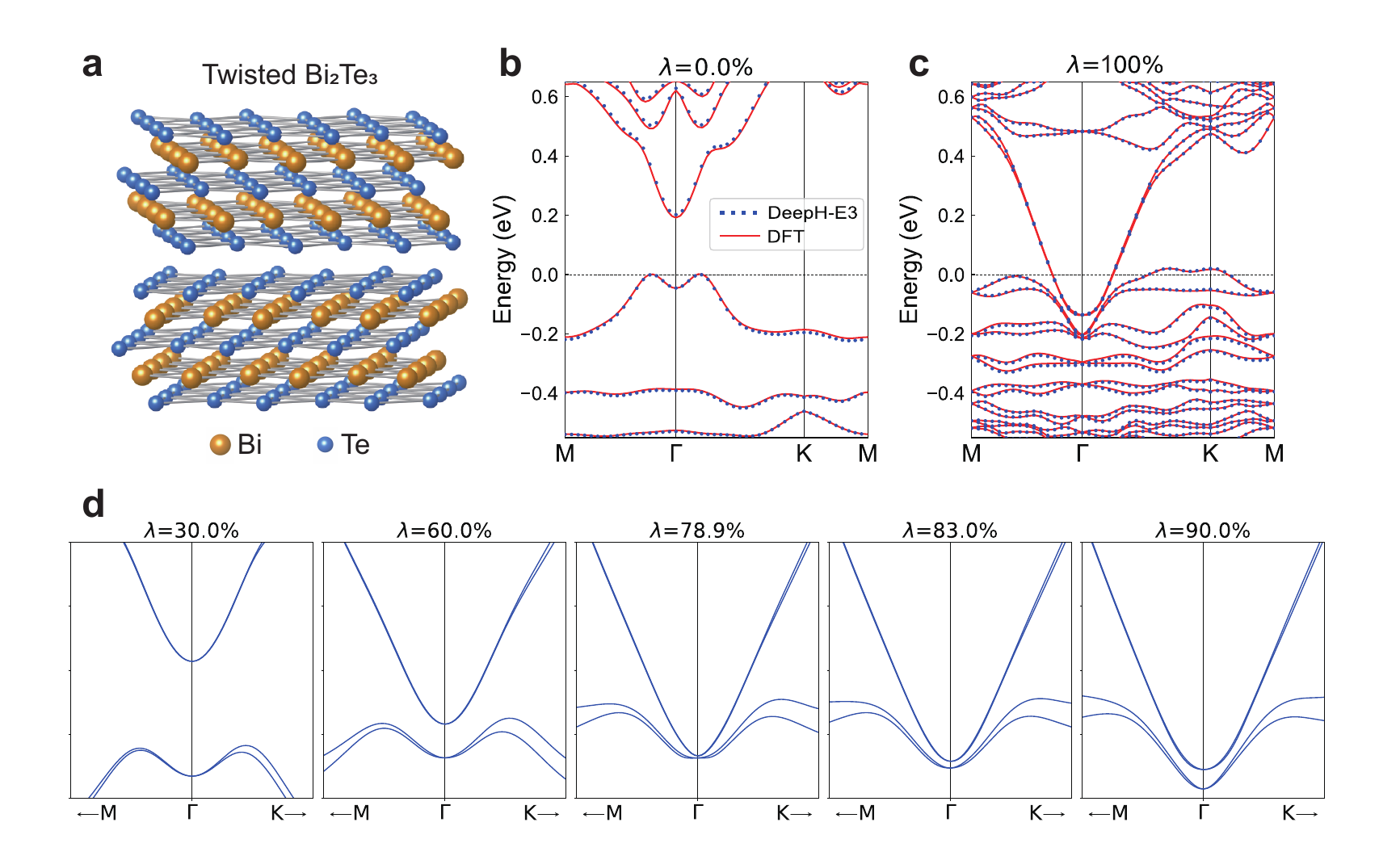}
    \caption{Application of DeepH-E3 to study SOC effects of twisted vdW materials. (\textbf{a}) Schematic structure of twisted bilayer Bi$_2$Te$_3$. (\textbf{b}, \textbf{c}) Band structures of bilayer Bi$_2$Te$_3$ with twist angle $\theta=21.8^\circ$ predicted by DeepH-E3 without SOC ($\lambda=0$) and with full SOC ($\lambda=1$), compared to DFT results. (\textbf{d}) Evolution of band structure as a function of SOC strength predicted by DeepH-E3. A closing and reopening of band gap at the $\Gamma$ point is visualized, indicating a topological quantum phase transition from $Z_2 = 0$ to $Z_2 = 1$ driven by SOC.}
    \label{fig:4.4-gapclose.pdf}
\end{figure*}

Furthermore, we have tested the performance of DeepH-E3 on studying twisted vdW materials with strong SOC, including twised bilayers of bismuthene, Bi$_2$Se$_3$ and Bi$_2$Te$_3$. The latter two materials are more complicated, which includes two quintuple layers and two kinds of elements (Fig.~\ref{fig:4.4-gapclose.pdf}a for Bi$_2$Te$_3$). The strong SOC introduces additional complexity in their electronic structure problems. Despite all these difficulties, the capability of DeepH-E3 is not influenced at any extent. Our method reaches sub-meV accuracy in predicting DFT Hamiltonian of test material samples, including untwisted and twisted structures of bismuthene, Bi$_2$Se$_3$ and Bi$_2$Te$_3$ bilayers. Impressively, the band structures predicted by DeepH-E3 match well with those obtained from DFT (supplementary Fig. S2). Moreover, we observe remarkable ability of our model to fit a tremendous amount of data with moderate model capacity and relatively small computational complexity. For instance, the neural network model is able to fit $2.8\times 10^9$ nonzero complex-valued Hamiltonian matrix elements in the dataset with about $10^5$ real parameters. The training time is about one day on a single GPU in order to reach sub-meV accuracy. More details are presented in the Supplementary Note 1. Through these experiments, the capability of DeepH-E3 on representing the spin-orbital DFT Hamiltonian is well demonstrated.

In physics, the SOC can induce many exotic quantum phenomena, leading to emergent research fields of spintronics, unconventional superconductivity, topological states of matter, etc. Investigation of SOC effects is thus of fundamental importance to the research of condensed matter physics and materials science. The functionality of analyzing SOC effects is easily implemented by DeepH-E3. Specifically, we apply two neural network models to learn DFT Hamitonians with full SOC ($\hat H_\text{1}$) and without SOC ($\hat H_\text{0}$) separately for the same material system. Then, we define a virtual Hamiltonian as a function of SOC strength ($\lambda$): $\hat H_{\lambda} = \hat H_{0} + \lambda \hat H_\text{SOC}$, where $\hat H_\text{SOC} = \hat H_\text{1} - \hat H_\text{0}$. By studying the virtual Hamiltonian at different $\lambda$, we can systematically analyze the influence of SOC effects on material properties.

As an example application, we employ the approach to investigate topological properties of twisted bilayer Bi$_{2}$Te$_{3}$. DeepH-E3 can accurately predict the DFT Hamitlonian for both cases with or without SOC, as confirmed by band-structure calculations using the predicted $\hat H_\text{DFT}$ (Figs.~\ref{fig:4.4-gapclose.pdf}b and \ref{fig:4.4-gapclose.pdf}c). Herein the SOC is extremely strong as caused by the heavy elements in the material. Consequently, the band structure changes considerably when SOC is turned on. The evolution of band structure as a function of SOC strength (Fig.~\ref{fig:4.4-gapclose.pdf}d) provides rich information on the SOC effects. Importantly, the band gap closes and reopens when increasing the  SOC strength, indicating a topological quantum phase transition from $Z_2 = 0$ to $Z_2 = 1$. This is further confirmed by applying symmetry indicators based on Kohn-Sham orbital analysis and by performing Brillouin-zone integration of Berry connection and curvature over all occupied states via the Fukui-Hatsugai-Suzuki formalism~\cite{Takahiro_2005}. The topological invariant $Z_2$ turns out to be nonzero for the spin-orbital coupled system, suggesting that the twisted bilayer Bi$_{2}$Te$_{3}$ ($\theta=21.8^\circ$) is topologically nontrivial. As DeepH-E3 works well for varying twist angle, the dependence of band topology on twist angle can be systematically computed, which will enrich the research of twisted vdW materials.

\section{DISCUSSION}

Since the DFT Hamiltonian $\hat H_\text{DFT}$ transforms covariantly between reference frames, it is natural and advantageous to construct the mapping from crystal structure $\{\mathcal R\}$ to $\hat H_\text{DFT}$ in an explicitly equivariant manner. In this context, we have developed a general framework to represent $\hat H_\text{DFT}$ with a deep neural network DeepH-E3 that fully respects the principle of covariance even in the presence of SOC. We have presented the theoretical basis, code implementation, and practical applications of DeepH-E3. The method enables accurate and efficient electronic structure calculation of large-scale material systems beyond the scope of traditional \textit{ab initio} approaches, opening possibilities to investigate rich physics and novel material properties at unprecedentedly low computational cost.

However, as the structure become larger, it becomes increasingly difficult to diagonalize the Hamiltonian matrix in order to obtain wavefunction-related physical quantities. This difficulty, instead of the limitations of DeepH-E3 method itself, will eventually become the bottleneck of accurate electronic structure predictions. Nevertheless, benefiting from the sparseness of the DFT Hamiltonian matrix under localized atomic-orbital basis, many efficient O($N$) algorithms with high parallel efficiency are available for studying large-scale systems (e.g., supercells including up to $10^7$ atoms~\cite{Hoshi2012}). Combination of the DeepH-E3 method with such efficient linear algebra algorithms will be a promising direction for future study.

The unique abilities of DeepH-E3 together with the general framework of incorporating symmetry requirements and physical insights into neural-network model design might find wide applications in various directions. For example, the method can be applied to build material database for a diverse family of Moir\'e-twisted materials. For each kind of material, only one trained neural network model will be needed for all the twisted structures in order to have full access to their electronic properties, which is a great advantage for high throughput material discovery. Moreover, since the deep-learning method does not rely on periodic boundary conditions, 2D materials with incommensurate twist angles can also be investigated, making \textit{ab initio} study of quasi-crystal phases possible. In addition, we could go one step further by calculating the derivative of the electronic Hamiltonian with respect to atomic positions via automatic differentiation techniques. This enables deep-learning investigation of the physics of electron-phonon coupling in large-scale materials, which has the potential to outperform the computationally expensive traditional methods of frozen phonon or density functional perturbation theory~\cite{Giustino2017}. Furthermore, one may combine the deep learning method with advanced methods beyond the DFT level, such as hybrid functionals, many-body perturbation theory, time-dependent DFT, etc. These important generalizations, if any of them are realized, would greatly enlarge the research scope of \textit{ab initio} calculation.

\appendix

\section{Datasets}

The monolayer graphene and MoS$_2$ datasets are taken from Ref.~\cite{deeph2022}. The datasets were generated by \textit{ab initio} molecular dynamics performed by the Vienna ab initio simulation package (VASP)~\cite{vasp1996}, using the projector-augmented wave pseudopotentials~\cite{paw1994,pawvasp1999} and the PBE~\cite{pbe} exchange-correlation functional. For the graphene dataset with 6$\times$6 supercells, 5000 frames were obtained at 300K with time step 1fs, then one frame was taken out every 10 frames starting from the 500th frame, thus there are 450 structures in the dataset. Among them, 270 were used for training, 90 for validation and 90 for testing. For the MoS$_2$ dataset with 5$\times$5 supercells, 1000 frames were taken at 300K with time step 1fs, and the first 500 unequilibrated structures were discarded. 300, 100, 100 structures were taken from the remaining 500 frames and used for training, validating and testing, respectively. The Hamiltonians of these structures were calculated using the OpenMX code~\cite{openmx2003, openmx2004} with the PBE exchange-correlation functional and the norm-conserving pseudopotentials distributed alongside the OpenMX package. The localized pseudo-atomic orbital basis used are C6.0-$s2p2d1$, Mo7.0-$s3p2d2$ and S7.0-$s2p2d1$, where the number behind the chemical symbol is the cutoff radius in a.u., $s2p2d1$ means there are $2\times1=2$ $s$-orbitals, $2\times3=6$ $p$-orbitals and $1\times5=5$ $d$-orbitals. The non-twisted bilayer graphene dataset is also taken from Ref.~\cite{deeph2022}. Altogether 300 structures with $4\times4$ supercells were generated. The sizes of training, validation and test sets are 180, 60, and 60, respectively. Non-twisted bilayer bismuthene, Bi$_2$Se$_3$ and Bi$_2$Te$_3$ datasets are all generated using a similar method in Ref.~\cite{deeph2022}. The bilayer unit cells of the latter two materials are relaxed using VASP~\cite{vasp1996} with vdW interaction corrected by DFT-D3 method with Becke-Jonson damping~\cite{BJ2005}. Supercells of $3\times3$ are prepared with uniform shift of interlayer stacking and random coordinate perturbation on each ions (within 0.1 \AA~along three directions), giving totally 576 structures. The localized pseudo-atomic orbital basis of Bi8.0-$s3p2d2$, Se7.0-$s3p2d1$ and Te7.0-$s3p2d2$ are used. The sizes of training, validation and test sets are 231, 113, 113, respectively, for bismuthene and Bi$_2$Se$_3$, and 346, 115, 115 for Bi$_2$Te$_3$.

\section{Details of neural network models}

All the neural network models presented in this article are trained by directly minimizing the mean-squared errors of the model output compared to the Hamiltonian matrices computed by DFT packages, and the reported MAEs are also obtained from comparing model output to the DFT results. All physical quantities of materials are derived from the output Hamiltonian matrix.

Some details of neural network building blocks are described here. The Gaussian basis is adopted from Ref.~\cite{schnet2018}, which is defined as:
\begin{equation}
    \boldsymbol e_\text{B}(|\boldsymbol r_{ij}|)_n= \exp \left(-\frac{\left(\left|\boldsymbol r_{i j}\right|-r_{n}\right)^{2}}{2\Delta^{2}}\right), \label{eq:gaussian-basis}
\end{equation}
where $r_n,n=0,1,\dots$ are evenly spaced, with intervals equal to $\Delta$. The E3Linear layer is defined as:
\begin{equation}
    \operatorname{E3Linear}(x)^{(l)}_{cm} = \sum_{c'=1}^{n^{(l)}}W^{(l)}_{cc'}x^{(l)}_{c'm}+b^{(l)}_{c},
\label{eq:E3Linear}
\end{equation}
where $W^{(l)}_{cc'},b^{(l)}_c$ are learnable weights and biases, $b^{(l)}_c=0$ for $l\neq 0$. In the gate layer, the $l=0$ part of the input feature is separated into two parts, denoted as $x_{1c}^{(0)}$ and $x_{2c}^{(0)}$. Notice that for $l=0$ so the index $m$ is omitted. The output feature is calculated by
\begin{equation}
    \operatorname{Gate}(x)^{(l)}_{cm} = 
    \begin{cases}
        \phi_1(x_{1c}^{(0)}), & l=0\\
    \phi_2(x_{2c}^{(0)})x^{(l)}_{cm}, & l\neq 0
    \end{cases}.
    \label{eq:gate-nonlin}
\end{equation}
Here $\phi_1$ and $\phi_2$ are activation functions. In this work, we use $\phi_1$=SiLU and $\phi_2$=Sigmoid following Ref.~\cite{nequip2022}.

The ENN is implemented with the e3nn library~\cite{e3nn} in version 0.3.5 and PyTorch~\cite{pytorch} in version 1.9.0. Our neural network consists of 3 vertex update blocks and 3 edge update blocks. The representation of the initial edge and vertex features are both $64\times 0\text{e}$. The intermediate edge and vertex features are $64\times0\text{e}+32\times1\text{o}+16\times2\text{e}+8\times3\text{o}+8\times4\text{e}+4\times5\text{o}$. Here $32\times1\text{o}$ denotes 32 vectors carrying the $l=1$ representation with odd parity, $16\times2\text{e}$ denotes 16 vectors carrying the $l=2$ representation with even parity. Spherical harmonics with $l=0$ to 5 are used. The Gaussian basis expansion used as input to the EquiConv layer has a length of 128. The fully connected neural network in the EquiConv layer is composed of two hidden layers, each with 64 hidden neurons, using the SiLU function as nonlinear activation and a linear layer as output. The batch size during all trainings is always set to 1. The initial learning rate is 0.003 for monolayer graphene and bilayer graphene, 0.005 for monolayer MoS$_2$, bilayer Bi$_2$Se$_3$ and bilayer bismuthene, and 0.004 for bilayer Bi$_2$Te$_3$. Learning rate is decreased by a factor of 0.5 when the loss plateaus, and the training is terminated when significant improvement cannot be reached. For the model with reduced number of parameters mentioned in the example study on Bi$_2$Se$_3$, the edge and vertex features used $32\times 0\text{e}+16\times 1\text{o}+8\times 2\text{e}+4\times 3\text{o}+4\times 4\text{e}$ and spherical harmonics has maximum $l=4$.

\begin{acknowledgments}
This work was supported by the Basic Science Center Project of NSFC (grant no. 51788104), the National Science Fund for Distinguished Young Scholars (grant no. 12025405), the National Natural Science Foundation of China (grant no. 11874035), the Ministry of Science and Technology of China (grant nos. 2018YFA0307100 and 2018YFA0305603), the Beijing Advanced Innovation Center for Future Chip (ICFC), and the Beijing Advanced Innovation Center for Materials Genome Engineering. R.X. was funded by the China Postdoctoral Science Foundation (grant no. 2021TQ0187).

X.G. and H.L. contributed equally to this work.
\end{acknowledgments}

\end{document}